\documentclass[12pt]{article}

\usepackage{amsmath}

\def\eq#1{Eq.~(\ref{#1})}
\newcommand{\secn}[1]{Section~\ref{#1}}

\newcommand{\nl}{\nonumber \\}

\def\beq{\begin{equation}}
\def\eeq{\end{equation}}
\def\beqa{\begin{eqnarray}}
\def\eeqa{\end{eqnarray}}
\def\ifm{\ifmmode}
\def\msb{\ifm \overline{\rm MS}\,\, \else $\overline{\rm MS}\,\, $\fi}

\def \fracs #1#2 {\mbox{\small $\frac{#1}{#2}$}}

\def \bin #1#2 {{\left({#1}\atop{#2}\right)}}
\def\lapproxeq{{\ \lower 0.6ex \hbox{$\buildrel<\over\sim$}\ }}
\def\gapproxeq{{\ \lower 0.6ex \hbox{$\buildrel>\over\sim$}\ }}

\def \as {\relax\ifmmode\alpha_s\else{$\alpha_s${ }}\fi}

\def \al #1 {\frac {\as({#1})}{\pi} }
\def \ds #1 {\ooalign{$\hfil/\hfil$\crcr$#1$}}

\def \a{\alpha}

\def\eq#1{Eq.~(\ref{#1})}
\def\gtil{{\widetilde {G}}}

\begin{document}

\begin{titlepage}

\rightline{DFTT-19/2008}
\rightline{NIKHEF/2008-008}
\rightline{ITP-UU-08/43}
\rightline{ITFA-2008-27}


\vspace{1.5cm}

\centerline{\Large \bf On next-to-eikonal corrections to threshold resummation}
\vspace{2mm}
\centerline{\Large \bf for the Drell-Yan and DIS cross sections} 

\vspace{1.2cm}

\centerline{\bf Eric Laenen\footnote{e-mail: {\tt Eric.Laenen@nikhef.nl}}}
\centerline{\sl NIKHEF Kruislaan 409, 1098 SJ Amsterdam}
\centerline{\sl ITFA, University of Amsterdam, Valckenierstraat 65, 1018 XE Amsterdam}
\centerline{\sl ITF, Utrecht University, Leuvenlaan 4, 3584 CE Utrecht, Netherlands}

\vspace{4mm}

\centerline{\bf Lorenzo Magnea\footnote{e-mail: {\tt magnea@to.infn.it}}}
\centerline{\sl Dipartimento di Fisica Teorica, Universit\`a di Torino}
\centerline{\sl and INFN, Sezione di Torino}
\centerline{\sl Via P. Giuria 1, I--10125 Torino, Italy}

\vspace{4mm}

\centerline{\bf Gerben Stavenga\footnote{e-mail: {\tt g.c.stavenga@uu.nl}}}
\centerline{\sl Institute for Theoretical Physics, Utrecht University}
\centerline{\sl Leuvenlaan 4, 3584 CE Utrecht, Netherlands}

\vspace{1.2cm}
 
\begin{abstract}

\noindent We study corrections suppressed by one power of the soft gluon
energy to the resummation of threshold logarithms for the Drell-Yan
cross section and for Deep Inelastic structure functions. While no
general factorization theorem is known for these next-to-eikonal (NE)
corrections, it is conjectured that at least a subset will
exponentiate, along with the logarithms arising at leading power. Here
we develop some general tools to study NE logarithms, and we construct
an ansatz for threshold resummation that includes various sources of
NE corrections, implementing in this context the improved collinear
evolution recently proposed by Dokshitzer, Marchesini and Salam (DMS).
We compare our ansatz to existing exact results at two and three
loops, finding evidence for the exponentiation of leading NE
logarithms and confirming the predictivity of DMS evolution.

\end{abstract}

\end{titlepage}

\newpage

\section{Introduction}
\label{intro}

Sudakov resummations are established in perturbative QCD for 
all logarithmic contributions, to leading power in the total momentum 
fraction carried by soft gluons. To illustrate this fact, consider as an 
example threshold resummation for the Drell-Yan process, or for a similar
electroweak annihilation cross section at the hard scale $Q$. In this case,
large logarithms arise in the hard partonic cross section when the total 
available center-of-mass energy, $\hat{s}$, is only slightly larger than
the mass $Q^2$ of the selected electroweak final state. Gluon radiation 
into the final state is then forced to be soft, as gluons carry (at most)
a total energy $(1 - z) \hat{s}$, with $z \equiv Q^2/\hat{s}$. As a 
consequence, perturbative contributions at order $\as^n$ are enhanced 
by large logarithms in the form of `plus' distributions, up to 
$\left[\ln^{2 n - 1} (1 - z)/(1 - z)\right]_+$. Upon taking a Mellin 
transform, these distributions turn into powers of logarithms of the Mellin 
variable $N$, conjugate to $z$, up to $\ln^{2 n} N$. All these contributions
can be resummed~\cite{Sterman:1986aj,Catani:1989ne}, and they display a
nontrivial pattern of exponentiation: the logarithm of the cross section
in Mellin space, in fact, is enhanced only by single logarithms, up to
$\ln^{n + 1} N$ at order $\as^n$.

It has been understood since the early days of QCD~\cite{Parisi:1979xd}
that at least some non-logarithmic contributions (terms independent of 
$N$, which are Mellin conjugate to virtual corrections proportional to 
$\delta(1 - z)$) also exponentiate. In fact, Ref.~\cite{Eynck:2003fn} 
later proved, at least for electroweak annihilation and DIS, that all such 
contributions can be organized in exponential form. One may naturally 
wonder to what extent this pattern of exponentiation can be extended 
beyond leading power in the Mellin variable $N$, or in the soft gluon 
energy fraction $1 - z$.

There are several problems in attempting to extend the resummation 
formalism beyond leading power in $N$, or $1 - z$. Indeed, resummation
can be understood to be a consequence of Sudakov factorization, as 
discussed in~\cite{Contopanagos:1996nh}. To leading power in $N$, it can 
be shown that the Mellin moments of the cross section factorize into distinct
functions responsible for infrared and collinear enhancements, times a hard
remainder which is free of logarithms. Exponentiation follows from evolution equations that are dictated by this factorization. To date, no proof of such a
Sudakov factorization is available beyond leading power in $N$. Part of this
issue is the fact that, in order the achieve exponentiation, the phase 
space specific to the observable at hand, in the threshold limit, must itself 
factorize; this is achieved at leading power by taking the Mellin transform,
thanks to the fact that the observable (essentially $1 - z$ for the inclusive 
Drell-Yan cross section) is linear in soft gluon energies to leading power in 
$1 - z$. Again, this simple property is lost beyond leading power.

Not withstanding these obstacles, there is intriguing, if scattered, 
evidence that some of the mechanisms that lead to the resummation 
of leading power logarithms are still operating at next-to-leading power.
Theoretically, evidence in this direction is provided by the Low-Burnett-Kroll
theorem~\cite{Low:1958sn,Burnett:1967km}, which states that (in QED) 
cross sections involving soft photons can be expressed in terms of 
radiation-less amplitudes not only at leading power in soft gluon energies 
(which corresponds to the bremsstrahlung spectrum and to the eikonal
approximation), but also at next-to-leading power. For such cross sections
radiation is simply related to classical fields, and one expects some form of 
soft photon exponentiation to hold. In QCD, direct application of Low's theorem 
is complicated by the presence of collinear divergences~\cite{DelDuca:1990gz},
but one may still expect it to be relevant for soft emissions.

At a more practical level, one may observe that resummed cross sections 
are expressed in terms of integrals of certain anomalous dimensions, with
integration limits dictated by the phase space available for soft radiation, and 
with the running coupling evaluated at the typical transverse momentum of 
the first gluon emission. These kinematical quantities are evaluated in the 
threshold limit, and one may expect that correcting their values in order 
to make them accurate at next-to-leading power in the soft momentum
should lead to a physically meaningful improvement of the resummation.

This kind of reasoning has led to attempts to include certain sub-eikonal effects
in practical implementations of Sudakov resummations, mostly in view of 
gauging the theoretical uncertainty of the resummation~\cite{Kramer:1996iq}.
Typically, this involves including subleading terms in the collinear evolution 
kernel into the resummation, which is particularly appealing for Drell-Yan and 
related cross sections, where the entire singularity structure is determined
by initial state soft and collinear radiation. This was applied in the case
of Higgs production in Refs.~\cite{Kramer:1996iq,Catani:2001ic,Harlander:2001is,Catani:2003zt},
and for prompt photon production in Ref.~\cite{Basu:2007nu}.

More recently, following the evaluation of collinear evolution kernels at three 
loops~\cite{Moch:2004pa}, a bold suggestion has been put forward by Dokshitzer, Marchesini and Salam (DMS)~\cite{Dokshitzer:2005bf}, who proposed a modified
evolution equation for parton distributions, based on the idea that the proper ordering variable in the collinear shower should be the lifetime of parton 
fluctuations rather than the gluon transverse momentum. This modified 
evolution has remarkable consequences: it explains a previously mysterious
numerical coincidence observed by~\cite{Moch:2004pa}, and it connects eikonal
and sub-eikonal terms in the splitting function in a nontrivial way, consistent 
with the idea that all evolution effects which are non-vanishing as $z \to 1$
should be determined at one loop, with an appropriate definiton of the 
coupling. The DMS proposal has later been refined by Basso and 
Korchemsky~\cite{Basso:2006nk}, who traced the recursive relation which
determines the collinear anomalous dimension to the conformal invariance
of the classical theory, and its breaking by the $\beta$ function. The relations
connecting eikonal and next-to-eikonal terms for parton evolution are then 
generalized to higher twist operators as well.

In this note, we begin to develop a systematic approach for the inclusion
of next-to-eikonal terms in the resummation, inspired by the results 
of~\cite{Dokshitzer:2005bf} and by the earlier work of~\cite{Kramer:1996iq}.
We begin, in \secn{tools}, by briefly reviewing the DMS approach, and 
describing how we intend to implement it in the context of Sudakov 
resummation. There, we also introduce some simple tools and definitions 
to evaluate the integrals that appear in resummed exponents to the desired
accuracy. Then, in \secn{subei}, we propose an ansatz to include in the
resummation all next-to-eikonal effects that can be argued to be under
theoretical control. We do this for the Drell-Yan cross section and for the 
Deep Inelastic structure function $F_2$. It is clear from the outset that our 
ansatz controls only a subset of all next-to-eikonal terms in the cross 
section: indeed, it may well be that not all such terms can be organized in exponential form. We believe however that the terms we include are physically 
well motivated, so we expect our ansatz to reproduce with reasonable accuracy
higher order perturbative results, based on the evaluation of the exponent at 
lower orders. We proceed to test this expectation by comparing the results
of expanding our proposed resummed expressions with the known exact results
at two loops for the Drell-Yan cross section~\cite{Hamberg:1990np}, and at two 
and three loops for DIS~\cite{Zijlstra:1992qd, Vermaseren:2005qc}.

As we will outline in our discussion, in \secn{discu}, the results of this 
comparison are consistent with the assumption that at least leading
next-to-eikonal logarithms do exponentiate, for all color structures. 
Furthermore, the implementation of the DMS approach reproduces
with considerable accuracy (though not exactly) certain classes of 
subleading next-to-eikonal logarithms which could not have been 
generated by the standard resummation. We believe that these results 
are encouraging regarding the possibility that next-to-eikonal logarithms 
could be understood and organized to all orders, an effort which will 
ultimately require a full analysis of soft gluon effects beyond the 
eikonal approximation.

\section{Tools for next-to-eikonal resummation}
\label{tools}

The task of probing  the extension of  the resummation formalism beyond 
the eikonal approximation requires both conceptual and practical tools. In this section we describe briefly the main conceptual progress that we are going to employ, which is the idea, put forward by DMS, that all NE terms in collinear
evolution trace their origin to one loop effects, phase space, and the choice 
of an appropriate, physically motivated coupling. We present the DMS equation, 
and we show how it can be solved in exponential form, just like ordinary 
collinear evolution, to NE accuracy. Next, making use of a technique 
developed in~\cite{Forte:2002ni}, which we generalize to NE level, we present
some simple results for the generic integrals that may appear in NE resummed
cross section to any perturbative order.

\subsection{The DMS evolution equation and its solution}
\label{dmsev}

Consider first the familiar collinear evolution equation for the non-singlet 
quark density
\beq
  \mu^2 \frac{\partial}{\partial \mu^2} \, q (x, \mu^2) = 
  \int_x^1  \frac{dz}{z} \,  q \left( \frac{x}{z}, \mu^2 \right) \, 
  P_{qq} \left( z, \as(\mu^2) \right) \, .
\label{dglap}
\eeq
As is well known, this simple convolution can be turned into a product by 
taking a Mellin transform,
\beq
  \mu^2 \frac{\partial}{\partial \mu^2} \, \tilde{q} (N, \mu^2) = 
  \gamma_N \left( \as(\mu^2) \right) \, \tilde{q} (N, \mu^2) \, ,
\label{dglapmom}
\eeq
which leads to an exponential solution for the Mellin moments of the quark distribution,
\beq
  \tilde{q} (N, \mu^2) = \exp \left[ \int_{\mu_0^2}^{\mu^2} 
  \frac{d \mu'^2}{\mu'^2} \gamma_N \left(\as(\mu'^2) \right) \right] \,
  \tilde{q} (N, \mu_0^2) \, .
\label{dglapsol}
\eeq
Note that here we express the solution in terms of a generic initial condition
at some reference scale, as appropriate for the evolution of physical, measured
parton distributions. When one instead considers parton-in-parton distributions,
defined in QCD in terms of matrix elements of bilocal operators, one can use dimensional regularization to express the solution as a pure exponential (with no
prefactor), using the fact that the dimensionally regularized coupling vanishes with 
the scale~\cite{Contopanagos:1996nh,Magnea:1990zb}. Within the framework
of dimensional regularization and in a minimal subtraction scheme, the structure
of the anomalous dimension $\gamma_N(\as)$ at large values of $N$ (corresponding to the $z \to 1$ limit) is known~\cite{Albino:2000cp} to be 
single-logarithmic. It is of the form
\beq
  \gamma_N \left( \as \right) = - A \left( \as \right) \ln \bar{N} + 
  B_\delta \left( \as \right) - C_\gamma \left( \as \right) \, 
  \frac{\ln\bar{N}}{N} + D_\gamma \left( \as \right) \, 
  \frac1N + {\cal O} \left( \frac{1}{N^2} \right) \, 
\label{gengam}  
\eeq
where the function $A(\as)$ is one half of the cusp anomalous dimension 
$\gamma_K (\as)$, and $\bar{N} = N \, {\rm e}^{\gamma_E}$. 
The DMS proposal is that the functions $C_\gamma (\as)$ 
and $D_\gamma (\as)$ are not genuinely independent, but they can be 
derived from the knowledge of $A(\as)$. In turn, $A(\as)$ can be interpreted 
as a definition of the coupling in a suitable scheme, which has been 
variously described as `physical', or `bremsstrahlung', or `Monte Carlo' 
scheme~\cite{Catani:1990rr}. In order to implement this idea, DMS propose to 
replace \eq{dglap} with
\beq
  \mu^2 \frac{\partial}{\partial \mu^2} \, \psi (x, \mu^2) = 
  \int_x^1 \frac{dz}{z} \,  \psi \left( \frac{x}{z}, z^\sigma \mu^2 \right) 
  {\cal P} \left( z, \as \left( \frac{\mu^2}{z} \right) \right) \, .
\label{dms}
\eeq
Here we have denoted by $\psi (x, \mu^2)$ a distribution which can be 
understood either as a fragmentation function or as a parton distribution; 
the parameter $\sigma = \pm 1$ serves to distinguish the two cases:
$\sigma = +1$ for the space-like evolution of parton distributions, while 
$\sigma = -1$ for the time-like evolution of fragmentation functions.
DMS argue (and verify at two loops) that with \eq{dms} the evolution kernel 
is the same for  both kinematics. Furthermore, at least up to second order 
in $\alpha_s$, the kernel ${\cal P}$ has no contributions at order $(1 - z)^0$, 
so that it can be written as
\beq
  {\cal P} \left( z, \as \right) = 
  \frac{A \left( \as \right)}{(1 - z)_+} 
  + B_\delta \left( \as \right) \delta (1 - z) + {\cal O} \left( (1 - z) 
  \right) \, .
\label{dmsans}
\eeq
If one now chooses the cusp anomalous dimension (divided by the Casimir 
invariant of the appropriate representation, in this case $C_F$) as the definition
of the coupling, setting $A\left(\as (\mu^2) \right) = C_F \, \alpha_{\rm PH}
(\mu^2)$, one may conclude that all contributions to the evolution kernel
that do not vanish as $z \to 1$ appear at the first non-trivial order in this 
scheme.

In the physical scheme, writing ${\cal P} (z, \alpha_{\rm PH}) = {\cal P}_1 
(z) \, \alpha_{\rm PH}/\pi + {\cal O} (\alpha_{\rm PH}^2)$, it is easy to construct an exponential solution, analogous to \eq{dglapsol} but valid to NE
order, for the distribution $D$. Indeed one may write
\beq
  \mu^2 \frac{\partial}{\partial \mu^2} \, \psi (N, \mu^2) = \int_0^1 dz \, 
  z^{N - 1} \, {\cal P}_1 ( z )  \, \alpha_{\rm PH} \left( \frac{\mu^2}{z} 
  \right)  \psi (N, z^\sigma \mu^2) \, .
\label{dmsmom}
\eeq
The scale of the coupling can be shifted by using the $\beta$ function, as
\beqa
  \mu^2 \frac{\partial}{\partial \mu^2} \, \psi (N, \mu^2) & = & 
  \int_0^1 dz \, z^{N - 1} \, 
  {\cal P}_1 (z) \Bigg[ \frac{\alpha_{\rm PH}}{\pi} 
  \label{dmsapp} \\
  && + \,  (1 - z) \, \left( \beta(\alpha_{\rm PH}) - \sigma \, 
  \frac{\alpha_{\rm PH}}{\pi} 
  \, \mu^2 \, \frac{\partial}{\partial \mu^2} \right) \Bigg] \, 
  \psi (N, \mu^2) \, \nonumber .
\eeqa
One can now perform a Mellin transform, and introduce the anomalous dimensions
\beq
  \widehat{\gamma}_1  (N) = \int_0^1 dz \, z^{N - 1} \, {\cal P}_1 (z) 
  \, , \qquad
  \widehat{\gamma}_1 \, ' (N) = \int_0^1 dz \, z^{N - 1} (1 - z) \, 
  {\cal P}_1 (z) \, ,
\eeq
which clearly obey $\widehat{\gamma} \, ' (N) = \widehat{\gamma} (N) - \widehat{\gamma} (N + 1)$. One finds then
\beq
  \psi (N, \mu^2) = \exp \left[\int_{\mu_0^2}^{\mu^2} 
  \frac{d \mu^2}{\mu^2} \frac{\widehat{\gamma}_1 ( N ) \, 
  \left( \alpha_{\rm PH} (\mu^2)/\pi \right)
 + \widehat{\gamma}_1 \, ' ( N ) \, \beta(\alpha_{\rm PH} 
  (\mu^2))}{1 + \sigma \, \widehat{\gamma} \, ' ( N ) \, 
  \left( \alpha_{\rm PH}(\mu^2)/\pi \right)}
  \right] \psi (N, \mu_0^2) \, ,
\label{dmssol}
\eeq
which is valid up to corrections vanishing as $z \to 1$.

\subsection{Moment integrals to ${\cal O}(1/N)$}
\label{integ}

Let us now turn to the practical issue of evaluating the generic integrals 
appearing in the exponents of threshold resummations, to our required 
accuracy, {\it i.e.} including all correction of order $1/N$. To this accuracy
threshold-resummed partonic cross sections can be written as
\beq
  \ln \big[ \hat{\sigma} ( N ) \big] - H = \int_0^1 dz \, 
  \frac{z^{N - 1} - 1}{1 - z} \, f_1 \big[ \ln(1 - z) \big] + 
  \int_0^1 dz \, z^{N - 1} f_2 \big[ \ln(1 - z) \big] \, ,
\label{genres}
\eeq
where $H$ represents $N$-independent terms. 
Expanding the functions $f_i$ in powers of their argument, as
\beq
  f_i \big[ \ln(1 - z) \big] = \sum_{p = 0}^\infty f_i^{(p)} \ln^p (1 - z) \, ,
\label{genser}
\eeq
we can write
\beq
  \ln \big[ \hat{\sigma} ( N ) \big] - H = \sum_{p = 0}^\infty 
  \left[f_1^{(p)} \, {\cal D}_p (N) + f_2^{(p)} \, {\cal J}_p (N) \right] \, ,
\label{genint}
\eeq
in terms of the basic integrals
\beq
  {\cal D}_p (N) = \int_0^1 dz \, \frac{z^{N - 1} - 1}{1 - z} \, \ln^p (1 - z) 
  \, , \qquad
  {\cal J}_p (N) = \int_0^1 dz \, z^{N - 1} \, \ln^p(1-z) \, .
\label{intdef}
\eeq
In order to evaluate the integrals in \eq{intdef}, we follow~\cite{Forte:2002ni}
and introduce two generating functions, defined by
\beq
  G_{\cal D} (\lambda, N) \equiv \int_0^1 (z^{N - 1} - 1) (1 - z)^{\lambda 
  - 1} = \frac{\Gamma(N) \Gamma(\lambda)}{\Gamma(N + \lambda)} - 
  \frac{1}{\lambda} \, ,
\label{GD}
\eeq
and by
\beq
  G_{\cal J} (\lambda, N) \equiv \int_0^1 z^{N - 1}\, (1 - z)^\lambda
  = \frac{\Gamma(N) \Gamma(\lambda + 1)}{\Gamma(N + \lambda + 1)}
  = \frac{1}{N + \lambda} \left[ \lambda \, G_{\cal D} (\lambda, N) + 1 
  \right] \, ,
\label{GJ}
\eeq
>From these definitions, one sees that the integrals in \eq{intdef} are given by
\beq
  {\cal D}_p (N) = \left. \frac{\partial^p}{\partial \lambda^p} \, G_{\cal D}
  (\lambda, N) \right|_{\lambda = 0} \, , \qquad
  {\cal J}_p (N) = \left. \frac{\partial^p}{\partial \lambda^p} \, G_{\cal J}
  (\lambda, N) \right|_{\lambda = 0} \, .
\label{generint}
\eeq
In order to evaluate the integrals explicitly to $1/N$ accuracy, we only need
the first correction to Stirling's formula for the ${\cal D}$-type integrals,
\beq
  \Gamma(z) = {\rm e}^{- z} \, z^{z - 1/2} \, \sqrt{2 \pi} \left(1 + 
  \frac{1}{12 z} \right) \left(1 + {\cal O} \left( \frac{1}{z^2} \right) 
  \right) \,  ,
\label{stirling}
\eeq
leading to
\beq
  G_{\cal D} (\lambda, N) = \frac{1}{\lambda} \left[
  \frac{\Gamma(1 + \lambda)}{N^\lambda}
  \left(1 + \frac{\lambda(1 - \lambda)}{2N} \right) - 1 \right] \, ,
\label{GDapp}
\eeq
while for the ${\cal J}$-type integrals it suffices to take
\beq
  G_{\cal J} (\lambda, N) = \frac{\Gamma(1 + \lambda)}{N^{1 + 
  \lambda}} \, .
\label{GJapp}
\eeq
We note in passing that, to $1/N$ accuracy, there is a simple relation 
between the ${\cal J}$ and the ${\cal D}$ integrals; in fact
\beq
  {\cal J}_p (N) = - \frac{d}{d N} {\cal D}_p (N) + {\cal O} \left( 
  \frac{1}{N^2}\right) \, ,
\label{JdD}
\eeq
which follows from an identical relation between the generating functions, 
\beq
  G_{\cal J} (\lambda, N) = - \frac{d}{d N} G_{\cal D} (\lambda, N) 
  + {\cal O} \left( \frac{1}{N^2} \right) \, .
\label{GJdGD}
\eeq
A useful way to evaluate both sets of integrals in the large $N$ limit is to 
map them into simpler integrals, where the dependence on $N$ has been 
moved from the integrand to the upper limit of integration. This technique 
is well known~\cite{Catani:1989ne,Forte:2002ni}, and we extend it here to $1/N$ accuracy. Let the generating function of cutoff integrals be
\beq
  G_L(\lambda, N) \equiv \int_0^{1 - 1/N} dz  \, (1 - z)^{\lambda - 1} 
  = \frac{1- N^{- \lambda}}{\lambda} \, .
\label{GL}
\eeq
It is then easy to relate this function to the functions $G_{\cal D}$ and 
$G_{\cal J}$. Expanding \eq{GDapp} in powers of $\lambda$ one finds
\beq
  G_{\cal D} (\lambda, N) = - G_L (\lambda, N) + \sum_{k = 1}^\infty \, 
  \frac{\Gamma_k (N)}{k!} \, \lambda^{k - 1} \, \frac{1}{N^\lambda} \, ,
\label{GDtoGL}
\eeq
where
\beq
  \Gamma_k (N) = \frac{d^k}{d \lambda^k} \left[ \Gamma(1 +
  \lambda) \left(1 + \frac{\lambda(1 - \lambda)}{2N} \right) 
  \right]_{\lambda = 0} \, .
\label{defgamk} 
\eeq
This can be rewritten as
\beq
  G_{\cal D} (\lambda, N) = \sum_{k = 0}^\infty 
  \frac{\Gamma_k (N)}{k!} (- 1)^{k - 1} \frac{\partial^k}{\partial (\ln N)^k}
  G_L(\lambda, N) \, .
\label{GDsumdGL}
\eeq
Using \eq{GJdGD} one then immediately finds
\beq
  G_{\cal J} (\lambda, N) = \frac{1}{N} \sum_{k = 0}^\infty 
  \frac{\Gamma_k (N)}{k!} (- 1)^k
  \frac{\partial^{k + 1}}{\partial (\ln N)^{k + 1}} G_L(\lambda, N) \, .
\label{GJsumdGL}
\eeq
Eqs.~(\ref{GDsumdGL}) and (\ref{GJsumdGL}) can be used to evaluate 
directly the ${\cal D}$ and ${\cal J}$ integrals to the desired accuracy, 
and indeed we will make use of this explicit evaluation in \secn{subei}. One 
finds
\beqa
  {\cal D}_p & = & \frac{1}{p + 1} \, \sum_{k = 0}^{p + 1} 
  \Gamma_k (N) \, \binom{p + 1}{k} \, (- \ln N)^{p + 1 - k}
  + {\cal O} \left( \frac{\ln^m N}{N^2} \right) \, , \nl
  {\cal J}_p & = & \frac{1}{N} \, \sum_{k = 0}^p 
  \Gamma^{(k)} (1) \, \binom{p}{k} \, ( - \ln N)^{p - k}
  + {\cal O} \left( \frac{\ln^m N}{N^2} \right) \, ,
\label{finDJ}  
\eeqa
where $\Gamma^{(k)}$ is the $k$'th derivative of the Euler gamma
function.
On the other hand, one can use Eqs.~(\ref{GDsumdGL}) and (\ref{GJsumdGL}) 
to directly relate the logarithm of the cross section to a cutoff integral of the
same functions $f_1$ and $f_2$ appearing in \eq{genres}. This is useful when 
one needs to correctly account for running coupling effects to all orders, as 
done in~\cite{Catani:1989ne,Forte:2002ni}. To the present accuracy one can write
\beqa
  \ln \big[ \hat{\sigma} ( N ) \big] - H & = &
  \sum_{k = 0}^{\infty} \frac{\Gamma_k (N)}{k!} ( - 1)^{k - 1}
  \frac{\partial^k}{\partial (\ln N)^k}
  \int_0^{1 - 1/N} d z \, \, \frac{f_1 \big[ \ln(1 - z) \big]}{1 - z}
  \label{fingenres} \\
  & - & \hspace{-2mm} 
  \frac{1}{N} \, \sum_{k = 0}^\infty \frac{\Gamma^{(k)} (1)}{k!}
  ( - 1)^{k - 1} \frac{\partial^{k + 1}}{\partial (\ln N)^{k + 1}}
  \int_0^{1 - 1/N} \hspace{-3pt} 
  d z \, \, \frac{f_2 \big[\ln(1 - z) \big]}{1 - z} \, . \nonumber
\eeqa
We now move on to applying these tools to the concrete example of
threshold resummation for the Drell-Yan and DIS cross sections.

\section{An ansatz for next-to-eikonal logarithms}
\label{subei}

In order to include NE effects in threshold resummation formulas we propose
to modify the exponents in three ways. First of all, following DMS, we include
subleading corrections in the argument of the running coupling. Second, we
change the boundary of phase space accordingly. Third, and most relevant,
we interpret the leading-logarithm function $A(\as)$ as arising from collinear
evolution, and thus replace it with a NE generalization dictated by the DMS 
equation. This is done in the following way. While \eq{dms} cannot be 
diagonalized by means of a simple Mellin transform, it is however possible, 
as pointed out by DMS, to map the kernel ${\cal P}(z, \as)$ in \eq{dmsans}
back to the conventional evolution kernel, order by order in perturbation 
theory, if one explicitly performs the shifts in the arguments of \eq{dms} by 
the action of differential operators. Indeed, one may rewrite \eq{dms} as
\beq
  \mu^2 \frac{\partial}{\partial \mu^2} \, \psi (x, \mu^2) = 
  \int_x^1  \frac{dz}{z} \,  {\rm e}^{- \ln z \, \left( \beta(\as) 
  \frac{\partial}{\partial \as} -\sigma \frac{\partial}{\partial \ln \mu^2} \right)} \, 
  \psi \left( \frac{x}{z}, \mu^2 \right) \, 
  {\cal P} \left( z, \as(\mu^2) \right) \, ,
\label{shift}
\eeq
where one should note that dependence on the coupling is only through the
kernel ${\cal P}$, while explicit scale dependence arises only in the distribution 
$\psi$. Expanding the exponential and the kernel ${\cal P}$ in perturbation 
theory one is led to an equation which can be diagonalized order by order. 
When solved in this way, the DMS equation can be understood as a framework
to generate classes of higher-order contributions to collinear anomalous 
dimensions using low-order information. In this spirit, we will write conventional
resummation formulas, but we will generalize the collinear evolution function
$A(\as)$ by including all terms that are generated by the DMS equation. As 
we will see, this will lead to slightly different implementations for space-like
and time-like kinematics. Let us now consider our two examples in turn.

\subsection{The Drell-Yan cross section}
\label{drelly}

We first consider the Drell-Yan hard partonic cross section in the $\msb$
factorization scheme, denoted $\widehat{\omega} (N)$. We propose to generalize
the exponentiation of threshold corrections in the following way.
\beqa
  \ln \Big[ \widehat{\omega} (N) \Big] & = & 
  {\cal F}_{\rm DY} \left( \a_s (Q^2) \right) + 
  \int_0^1 \, dz \, z^{N - 1} \, \Bigg\{ \frac{1}{1 - z} \, 
  D \left[ \as \left( \frac{(1 - z)^2 Q^2}{z} \right) \right] \nonumber \\ & + &
  2 \,\int_{Q^2}^{(1 - z)^2 Q^2/z} \, 
  \frac{d q^2}{q^2} \, P_s \Big[ z, \as (q^2) \Big] \Bigg\}_+ \, ,
\label{newresDY}
\eeqa
where for simplicity we have set the factorization scale $\mu_F^2 = Q^2$.
Here and below we adopt the convention that the `plus' prescription applies
only to singular terms in the expansion of the relevant functions in powers
of $1 - z$. In other words, for a singular function $f(z)$ with Laurent expansion
$f(z) = \sum_{n = -1}^\infty f_n \, (1 - z)^n$, and for any smooth function 
$g(z)$, regular as $z \to 1$, we define
\beq
  \int_0^1 d z \, g (z) \Big[ f(z) \Big]_+ \equiv f_{-1} \int_0^1 d z \, 
  \frac{g(z) - g(1)}{1 - z} + \int_0^1 d z \, g(z) \left( f(z) - 
  \frac{f_{-1}}{1 - z} \right)~.
\label{plusdef}
\eeq
In \eq{newresDY}, ${\cal F}_{\rm DY} (\as)$ is responsible for the 
exponentiaton of $N$-independent terms, in accordance 
with~\cite{Eynck:2003fn}. It comprises purely virtual contributions
given in terms the quark form factor, and real emission terms, which were 
denoted by $F_{\msb} (\as)$ in~\cite{Eynck:2003fn}. The single-logarithm
function $D(\as)$ can also be related to form factor data, and to the virtual 
part of the collinear evolution kernel $B_\delta (\as)$, as was done 
in~\cite{Laenen:2005uz}, according to
\beq
 D(\alpha_s) = 4 \, B_\delta (\alpha_s) - 2 \, \gtil (\alpha_s) 
 + \beta (\alpha_s) \, \frac{d}{d \alpha_s} F_{\msb} (\alpha_s)~,
\label{D} 
\eeq
where $\gtil$ is constructed from single pole contributions to the quark form 
factor, as described in~\cite{Eynck:2003fn}. Finally, the DMS-improved space-like
collinear evolution kernel $P_s (z, \as)$ is given in perturbation theory by 
$P_s (z, \as) = \sum_{n = 1}^\infty P_s^{(n)} (z) \, (\as/\pi)^n$, where
\beq
  P_s^{(n)} (z) = \frac{z}{1 - z} A^{(n)} + C_\gamma^{(n)} \ln (1 - z) + 
  \overline{D}_\gamma ^{(n)}~.
\label{improP}
\eeq
Here $A^{(n)}$ and $C_\gamma^{(n)}$ are the perturbative coefficients of 
the functions appearing in \eq{gengam}, while $\overline{D}_\gamma^{(n)}$ 
is related to the perturbative coefficients of $D_\gamma (\as)$ by the simple
shift $\overline{D}_\gamma^{(n)} = D_\gamma^{(n)} + A^{(n)}$; this takes
into account the explicit factor of $z$ multiplying $A(\as)$ in \eq{improP},
which in turn is responsible for the inclusion of NE terms in the ordinary
evolution kernel. In our normalization, $A^{(1)} = C_F$, $C_\gamma^{(1)} =
\overline{D}_\gamma^{(1)} = 0$, while at two loops
\beqa
  A^{(2)} & = & \frac{1}{2} \left[ \left(\frac{67}{18} - \zeta(2) \right) 
  C_A C_F - \frac{5}{9} n_f C_F \right] \, , 
  \qquad C_\gamma^{(2)} = C_F^2 \, , \nonumber \\
  \overline{D}^{(2)} & = & \frac{3}{4} C_F^2 - \frac{11}{12} C_A C_F + 
  \frac{1}{6} n_f C_F \, .
\label{twolDYco}
\eeqa
Notice in particular that the DMS procedure has brought into the resummation
exponent abelian-like terms proportional to $C_F^2$ at two loops. As we will 
see, these terms do indeed find a match in the finite order expansion of 
$\widehat{\omega} (N)$. 
The ansatz (\ref{newresDY}) can be written in form of Eq.~(\ref{genres}),
and evaluated using the methods of \secn{tools}.
In \secn{discu}, we will compare the perturbative expansion of \eq{newresDY}, with the coefficients given in \eq{twolDYco}, to 
the exact results of~\cite{Hamberg:1990np}. 
In both cases, one may write the expansion 
\beq
  \widehat{\omega} (N) = \sum_{i = 0}^\infty \left( \frac{\as}{\pi} \right)^n 
  \Bigg[ \sum_{m = 0}^{2 n} \, a_{n m} \ln^m \bar{N} + 
  \sum_{m = 0}^{2 n - 1} b_{n m} \frac{\ln^m \bar{N}}{N} \Bigg] 
  + {\cal O} \left( \frac{\ln^p N}{N^2} \right) \, , 
\label{DYcoeff}
\eeq
and then compare the 
expressions for the coefficients $a_{nm}$ and $b_{nm}$ arising from 
the resummation to the exact ones.

\subsection{DIS structure functions}
\label{disf2}

We consider next the resummation for the DIS structure function 
$\widehat{F}_2 (N)$, in the $\msb$ factorization scheme. Phase 
space and kinematics in this case are somewhat more complicated, 
since one has to deal with the final state jet, which is approximately 
massless near threshold, as well as with initial state soft and collinear 
radiation. We propose to generalize the conventional resummation 
formula as
\beqa
  && \hspace{-8mm} \ln \Big[ \widehat{F}_2 (N) \Big] \, = \,  
  {\cal F}_{\rm DIS} \left( \a_s (Q^2) \right) + 
  \int_0^1 \, dz \, z^{N - 1} \, \Bigg\{ \frac{1}{1 - z} \, 
  B \left[ \as \left( \frac{(1 - z) Q^2}{z} \right) \right] \nonumber \\ & + &
  \int_{Q^2}^{(1 - z) Q^2/z} \, 
  \frac{d q^2}{q^2} \, P_s \Big[ z, \as (q^2) \Big]  \, + \, 
  \int_{(1 - z)^2 Q^2/z}^{(1 - z) Q^2/z} \, 
  \frac{d q^2}{q^2} \, \, \delta P \Big[ z, \as (q^2) \Big]
  \Bigg\}_+ \, .
\label{newresDIS}
\eeqa
Here, as above, ${\cal F}_{\rm DIS} ( \as)$ is responsible for the exponentiation 
of $N$-independent terms. The case of the DIS cross section in the $\msb$ factorization scheme was not explicitly treated in Ref.~\cite{Eynck:2003fn}, 
but it is easy to work out the relevant contributions from the information 
collected there. Indeed, one can reconstruct the structure function 
$\widehat{F}_2 (N)$ from the moment space ratio of the Drell-Yan cross 
section computed in the $\msb$ scheme to that computed 
in the DIS scheme, both given in \cite{Eynck:2003fn}, 
as $\widehat{F}_2^{(\msb)} (N) = \sqrt{
\widehat{\omega}^{(\msb)} (N)/\widehat{\omega}^{({\rm DIS})} (N)}$. One 
then easily verifies that ${\cal F}_{\rm DIS} ( \as)$ comprises a virtual part, 
given by the finite terms in the modulus squared of the space-like quark form factor, plus a combination of real emission contributions, which can be written 
as $\left(F_{\msb} (\as) - F_{\rm DIS} (\as) \right)/2$ in the notation 
of~\cite{Eynck:2003fn}. The single-logarithm function $B(\as)$ can be 
associated with the evolution of the final state jet. It is interesting to note
here that $B(\as)$ can also be expressed in terms of form factor data, plus
virtual corrections to the collinear evolution kernel, plus a total derivative
of lower order contributions, just like the function $D(\as)$ in \eq{D}. Indeed,
one verifies that existing results up to three loops are consistent with
\beq
 B (\alpha_s) =  B_\delta (\alpha_s)  -  \gtil (\alpha_s) 
 + \beta (\alpha_s) \, \frac{d}{d \alpha_s} F_{\rm B} (\alpha_s)~,
\label{B} 
\eeq
with easily computed perturbative coefficients for the function $F_{\rm B} 
(\as)$. \eq{B} is in keeping with the general results of Ref.~\cite{Dixon:2008gr},
where it was shown, at the amplitude level, that all IR and collinear singularities
in massless gauge theories can be constructed from combinations of eikonal functions with the virtual collinear function $B_\delta (\as)$, up to total
derivatives with respect to the scale. Finally, we turn to the second line of 
\eq{newresDIS}. There, we have used the fact that the integration over the 
scale $q^2$ has a range that can be split into two intervals, which correspond 
to different physical sources of radiation. Scales between the factorization scale $Q^2$ and the soft scale $(1 - z)^2 Q^2$ correspond to Drell-Yan-like initial
state radiation, while scales between the soft scale and the jet scale, 
$(1 - z) Q^2$, correspond to the evolution of the final state jet. Accordingly, in 
the first range we use the same space-like evolution kernel $P_s (z, \as)$ that
was employed in \eq{newresDY}, while in the second range we use the time-like 
fragmentation kernel $P_t (z, \as)$. One may then define $\delta P (z, \as)
\equiv P_t (z, \as) - P_s (z, \as)$, and thus get to \eq{newresDIS}. The 
function $\delta P (z, \as)$ begins at two loops, where it is given 
by~\cite{Curci:1980uw}
\beq
  \delta P^{(2)} (z) = - \frac{1}{2} \, C_F^2 \, \left (4 \ln (1 - z) + 
  3 \right)  + {\cal O} (1 - z) \, .
\label{curfur}
\eeq
Once again, using the methods of \secn{tools}, we can expand both the 
resummed and the exact results for $\widehat{F}_2 (N)$ in powers of 
logarithms of $\overline{N}$, and in inverse powers of $N$, as
\beq
  \widehat{F}_2 (N) = \sum_{i = 0}^\infty \left( \frac{\as}{\pi} \right)^n 
  \Bigg[ \sum_{m = 0}^{2 n} \, c_{n m} \ln^m \bar{N} + 
  \sum_{m = 0}^{2 n - 1} d_{n m} \frac{\ln^m \bar{N}}{N} \Bigg] 
  + {\cal O} \left( \frac{\ln^p N}{N^2} \right) \, .
\label{DIScoeff}
\eeq
We can then compare the resummed and exact values of the coefficients
$c_{nm}$ and $d_{nm}$, up to two and three loops, using the results 
of~\cite{Zijlstra:1992qd,Vermaseren:2005qc}.

\section{Discussion}
\label{discu}

We begin by checking the behavior of our ansatz at the one loop level. This 
is not trivial, since we have not added new coefficients in the exponent at 
one loop, and the only sources of $1/N$ terms are the expansions of the 
${\cal D}_p$ integrals, and the simple modifications of phase space. 
Using the one loop results for the functions $A(\as)$ and $D(\as)$, we 
find that for the Drell-Yan cross section the one-loop exact
result is recovered, including all corrections down to ${\cal O} (1/N)$. 
Specifically, expanding \eq{newresDY}, we find $b_{11} = 2 C_F$ and 
$b_{10} = 0$, which is exact. Note that $b_{10}$ vanishes as a 
consequence of a cancellation between subleading terms in the expansion
of the ${\cal D}_p$ integrals and the modified phase space boundary.
For DIS, including the one-loop value of the function $B(\as)$, we find 
that $d_{11} = C_F/2$ is correctly reproduced, while the non-logarithmic
term at ${\cal O} (1/N)$ is underestimated: \eq{newresDIS} yields 
$d_{10} = C_F/8$, while the exact result is $d_{10} = 21/8 \, C_F$. We take 
this as evidence (to be reinforced below) that our treatment of phase space
for the final state jet is sufficiently precise to reproduce single NE logarithms,
but not enough to fix NE constants (of course at this level non-factorizing 
effects for the observable, leading to a failure of exponentiation, at least in 
the form of \eq{newresDIS}, may also be a source of the discrepancy).

\begin{table}[hbt]

\begin{center}

\begin{tabular}{|c|c|c|c|c|c|c|}
 
 \hline
    & \multicolumn{2}{c|}{$C_F^2$}  &  \multicolumn{2}{c|}{$C_AC_F$}  
    & \multicolumn{2}{c|}{$n_fC_F$} \\ 
  \hline
  \hline
   $b_{23}$   &   $4$   &   $4$   &   $0$   &   $0$   &   $0$   &   $0$   \\
   $b_{22}$   & $ \frac{7}{2}$  & 4 & $\frac{11}{6}$ & $\frac{11}{6}$ & 
   $- \frac{1}{3}$ & $- \frac{1}{3}$  \\
   $b_{21}$   &   $8 \zeta_2 - \frac{43}{4}$   &   $8 \zeta_2 - 11$   &   
   $- \zeta_2 + \frac{239}{36} $   &   $- \zeta_2 + \frac{133}{18} $   &   
   $- \frac{11}{9}$   &   $- \frac{11}{9}$   \\
   $b_{20}$   &   $- \frac{1}{2} \zeta_2 - \frac{3}{4}$   &   $4 \zeta_2$   &   
   $- \frac{7}{4} \zeta_3 + \frac{275}{216}$   &   
   $\frac{7}{4} \zeta_3 + \frac{11}{3} \zeta_2 - \frac{101}{54}$   &   
   $- \frac{19}{27}$   &     $- \frac{2}{3} \zeta_2 + \frac{7}{27}$   \\
  \hline

\end{tabular}

\end{center}

\caption{\label{tab:dycoeff} 
Comparison of exact and resummed 2-loop coefficients for the Drell-Yan cross 
section. For each color structure, the left column contains the exact results, the right column contains the prediction from resummation.}

\end{table}

At the two-loop level, we proceed as follows. Since our aim is to verify our 
ability to reproduce NE terms, suppressed by a power of $N$, we include in 
the exponent all terms that are required to reproduce ordinary Sudakov
logarithms, {\it i.e.} the two-loop values of the functions $A(\as)$ and 
$D(\as)$ for the Drell-Yan cross section, and of the function $B(\as)$ for 
DIS. We include the two-loop DMS-induced contributions $C_\gamma^{(2)}$
$\overline{D}_\gamma^{(2)}$ and $\delta P^{(2)} (z)$ as well, since they 
are responsible for effects that originate at two loops, and can only be 
reproduced by their inclusion. Our results are summarized in Tables 1 (for
the Drell-Yan cross section) and in Table 2 (for the DIS structure function).

\noindent We observe the following.
\begin{itemize}

\item The leading non-vanishing NE logarithms ($\ln^3 \bar{N}/N$ for
the `abelian' terms proportional to $C_F^2$, and $\ln^2 \bar{N}/N$ for 
non-abelian terms) are correctly reproduced by the exponentiation, both for DY 
and for DIS, and separately for each color structure.

\item Next-to-leading NE logarithms ($\ln^2 \bar{N}/N$ for
terms proportional to $C_F^2$, and $\ln \bar{N}/N$ for 
non-abelian terms) are reproduced with remarkable accuracy for the 
Drell-Yan  process (in fact exactly for the $n_f C_F$ color structure), and
reasonably well for the DIS process.

\item The remaining NE logarithms, {\it i.e.} single logarithmic terms 
proportional to $C_F^2$, are well reproduced by exponentiation for the 
Drell-Yan process, but only roughly approximated for DIS. Non-logarithmic
NE corrections are not well approximated by the exponentiation.

\item More specifically, we note that for the Drell-Yan process the only 
source of terms proportional to $C_F^2 \ln^2 \bar{N}/N$ is the
DMS-induced coefficient $C_\gamma^{(2)}$; indeed, the fact that 
$b_{10} = 0$ ensures that no such term can arise from the square of
the one-loop contribution. This contribution, yielding $b_{22} = 4$, is an
excellent approximation to the exact result, $b_{22} = 7/2$. For DIS, 
as might be expected, the situation is somewhat more intricate; indeed
$d_{22}$ receives contributions from three sources: the square of the 
one-loop exponent, $C_\gamma^{(2)}$, and $\delta P^{(2)} (z)$; also
here, however, the final result, $d_{22} = 55/16$, is a fair approximation
of the exact answer, $d_{22} = 39/16$.

\end{itemize}

\begin{table}[hbt]

\begin{center}

\begin{tabular}{|c|c|c|c|c|c|c|}
 
 \hline
    & \multicolumn{2}{c|}{$C_F^2$}  &  \multicolumn{2}{c|}{$C_AC_F$}  
    & \multicolumn{2}{c|}{$n_fC_F$} \\ 
  \hline
  \hline
   $d_{23}$   &   $\frac{1}{4}$   &   $\frac{1}{4}$   &   $0$   &   $0$   &   
   $0$   &   $0$   \\
   $d_{22}$   & $ \frac{39}{16}$  & $\frac{55}{16}$ & $\frac{11}{48}$ & 
   $\frac{11}{48}$ & $- \frac{1}{24}$ & $- \frac{1}{24}$  \\
   $d_{21}$   &   $\frac{7}{4} \zeta_2   - \frac{49}{32}$   &   
   $- \frac{1}{4} \zeta_2 - \frac{105}{32}$   &   
   $- \frac{5}{4} \zeta_2 + \frac{1333}{288}$   &   
   $-  \frac{1}{4} \zeta_2 + \frac{565}{288}$   &   
   $- \frac{107}{144}$   &   $-  \frac{47}{144}$   \\
   $d_{20}$   &   $\frac{15}{4} \zeta_3 - \frac{47}{16} \zeta_2$   & 
   $- \frac{3}{4} \zeta_3 + \frac{53}{16} \zeta_2$ &  
   $- \frac{11}{4} \zeta_3 + \frac{13}{48} \zeta_2 $   &   
   $\frac{5}{4} \zeta_3 + \frac{7}{16} \zeta_2$   &   
   $\frac{1}{24} \zeta_2 - \frac{1699}{864}$   &   
   $- \frac{1}{8} \zeta_2 + \frac{73}{864}$   \\
   & $\hspace{5mm} - \, \frac{431}{64}$ & 
   $ \hspace{8mm} - \, \frac{21}{64}$ & 
   $ \hspace{6mm} - \, \frac{17579}{1728}$ &
   $ \hspace{6mm} - \, \frac{953}{1728}$
   & &  \\
  \hline

\end{tabular}

\end{center}

\caption{\label{tab:discoeff} 
Comparison of exact and resummed 2-loop coefficients for the DIS structure function. For each color structure, the left column contains the exact results, 
the right column contains the prediction from resummation.}

\end{table}

\noindent
Clearly, since some of the DMS modifications enter the stage at two-loops, 
our results verify that these contributions improve the approximation, but
do not really test exponentiation. We can put at least our DIS ansatz to a 
more stringent test by comparing to the complete three-loop calculation
performed by Moch, Vermaseren and Vogt~\cite{Vermaseren:2005qc}. 
In this case, since our aim is to test exponentiation 
at NE level, we have included the three-loop value of the function 
$B (\as)$, contributing to single Sudakov logarithms, but we have not 
included three-loop DMS-induced contributions such as $C_\gamma^{(3)}$ 
and $\delta P^{(3)} (z)$. We can then expect reasonable agreement 
only for a limited set of NE logarithms. Since at three loops
one finds six independent color structures, up to five powers of NE 
logarithms, and transcendentals up to $\zeta_5$, we do not include here
the lengthy tables of coefficients, but we give the most relevant results.

The three-loop analysis confirms that leading non-vanishing NE logarithms
(in this case $\ln^5 \bar{N}/N$ for the color structure $C_F^3$,
$\ln^4 \bar{N}/N$ for the color structures $C_A C_F^2$ and 
$n_f C_F^2$, and $\ln^3 \bar{N}/N$ for the color structures 
$C_A^2 C_F$, $n_f^2 C_F$ and $n_f C_A C_F$) are exactly reproduced 
by our resummation ansatz. Next-to-leading NE logarithms are reasonably 
well reproduced: specifically, for all color structures and separately for
each degree of transcendentality the approximate results from the 
resummation have the same sign and similar numerical values to the
corresponding exact results. In particular, this applies to the coefficient
$d_{34}$, whose exact value is $57/64$, while the approximate result
is $109/64$. Since $d_{34}$ arises in part from interference between
the NE coefficient $C_\gamma^{(2)}$ and the leading one-loop Sudakov
logarithms in the exponent, we take this as mild evidence in favor of the exponentiation of DMS-induced corrections.

To summarize, we have provided an ansatz to include in threshold 
resummation a set of next-to-eikonal corrections, allowing for subleading 
phase-space effects, and including the modified collinear evolution proposed 
by Dokshitzer, Marchesini and Salam. It is understood that these modifications
of conventional threshold resummation do not exhaust all possible sources
of NE threshold logarithms, and indeed it may be expected
that some such corrections might break Sudakov factorization and fail to exponentiate. By comparing our ansatz to finite order perturbative results
for the Drell-Yan and DIS cross sections, up to three loops, we have however
provided evidence that at least the leading non-vanishing NE logarithms do 
indeed exponentiate according to our proposal. We have furthermore provided
evidence that the DMS equation induces a definite improvement for
resummation at NE level: for example, abelian-like next-to-leading NE
terms that conventional resummation completely fails to generate are
accurately approximated when DMS evolution is implemented. In general, it 
is clear that our ansatz gives better results for the Drell-Yan process, 
presumably thanks to its simple phase space and kinematics.
The presence of the final state jet in DIS, and the related constraints on 
phase space, may require a more detailed factorization analysis in order
to collect all sources of NE terms, and indeed may well induce a breakdown
of simple Sudakov factorization at NE level. To aid this preliminary exploration 
of NE exponentiation, we have provided here some practical tools that will be
useful in future extensions of this work, and we have taken the opportunity to 
note a connection, given in \eq{B}, between the jet function $B(\as)$ and 
the virtual collinear function $B_\delta (\as)$, as was previously done for 
the soft function $D(\as)$ in the Drell-Yan cross section~\cite{Laenen:2005uz}.
We believe that this work provides further motivation both to include leading
NE correction in phenomenological resummation studies, and to pursue
the corresponding theoretical work. Indeed, a full understanding of NE 
threshold logarithms must await a thorough analysis of soft gluon radiation 
beyond the eikonal approximation in the non-abelian theory, and specifically 
an adequate implementation of Low's theorem, mapping its boundaries of
applicability in the case of massless QCD.

\vspace{2cm}

{\large{\bf Acknowledgements}}

\vspace{2mm}

\noindent We thank A. Vogt for providing us 
with code implementing three-loop DIS results in a form useful for our calculation.
We thank Chris White for several discussions concerning 
sub-eikonal corrections. L.M. thanks NIKHEF, and the CERN PH Department, 
TH Unit for hospitality during the completion of this work. E.L.  
thanks the University of Torino and INFN, Sezione di Torino for hospitality.
This work was supported in part by MIUR under contract 2006020509$\_$004, 
by the European Community's Marie-Curie Research Training Network 
`Tools and Precision Calculations for Physics Discoveries at Colliders' 
(`HEPTOOLS'), under contract MRTN-CT-2006-035505, by the Foundation 
for Fundamental Research of Matter (FOM), and by the National Organization 
for Scientific Research (NWO).

\end{document}